# Low-complexity Approaches for MIMO Capacity with Per-antenna Power Constraint

Thuy M. Pham, Ronan Farrell, and Le-Nam Tran
Department of Electronic Engineering, Maynooth University, Co. Kildare, Ireland
Email: {minhthuy.pham, ronan.farrell, lenam.tran}@nuim.ie

*Abstract*—This paper proposes two low-complexity iterative algorithms to compute the capacity of a single-user multiple-input multiple-output channel with per-antenna power constraint. The first method results from manipulating the optimality conditions of the considered problem and applying fixed-point iteration. In the second approach, we transform the considered problem into a minimax optimization program using the well-known MAC- BC duality, and then solve it by a novel alternating optimization method. In both proposed iterative methods, each iteration involves an optimization problem which can be efficiently solved by the water-filling algorithm. The proposed iterative methods are provably convergent. Complexity analysis and extensive numerical experiments are carried out to demonstrate the superior performance of the proposed algorithms over an existing approach known as the mode-dropping algorithm.

## I. INTRODUCTION

The studies on multiple-input multiple-output (MIMO) communications systems were pioneered by the seminal work of Foschini *et al.* [1] and Teletar [2], which characterized the capacity of a single-user MIMO with a sum power constraint. In reality, each transmit antenna is usually connected to an individual power amplifier (PA), and thus should be subject to its own power budget. The main purpose is to force these PAs to operate in the linear region to reduce non-linear distortion and to improve their efficiency. In this paper we consider the fundamental problem of MIMO capacity with per-antenna power constraint (PAPC)

Despite its importance, research on MIMO capacity for single user MIMO (SU-MIMO) subject to PAPC has been relatively sparse, especially viewed from a signal processing perspective. In fact, the problem of capacity computation of SU-MIMO with PAPC is known to be convex, and perhaps this recognition gives an impression that generic convex solvers or standard techniques such as interior-point methods are adequate. However, the complexity of such a method increases rapidly with the problem size, and thus is not appealing for massive MIMO systems. An effort was made in [3], [4] to to find the capacity of SU-MIMO with PAPC by closed-form expressions. However, the proposed algorithm in [3], [4] lacked a complete convergence proof. Moreover, the assumption of full-rank channel matrix is not likely met in reality.

In this paper, we proposed two fast-converging low-complexity iterative algorithms to compute the optimal input covariance matrices under PAPC. The first method is based on manipulating the optimality conditions of the considered problem and fixed-point iteration. The second one relies on the well-known MAC-BC duality but the resulting minimax problem is solved by a novel alternating optimization (AO) algorithm. Specifically, we proposed to optimize the upper bound of the objective with respect to a coordinate, eliminating the zigzag effect likely occurring in a pure AO method. Both proposed methods are provably convergent without any specific assumption on the channel matrix.

The remainder of the paper is organized as follows. The system model is described in Section II followed by the algorithm descriptions in Section III and IV. Section V provides the complexity analysis of the algorithms while Section VI presents the numerical results. Finally, we conclude the paper in Section VII.

*Notation*: Standard notations are used in this paper. Bold lower and upper case letters represent vectors and matrices, respectively; $\mathbf{H}^{\dagger}$ and $|\mathbf{H}|$ are the Hermitian and determinant of $\mathbf{H}$, respectively; $\mathrm{diag}(\mathbf{x})$ denotes a diagonal matrix with elements $\mathbf{x}$; $\mathrm{diag}(\mathbf{H})$, where $\mathbf{H}$ is a square matrix, returns the vector of diagonal elements of $\mathbf{H}$. The notation $\mathbf{x} \odot \mathbf{y}$ denotes the Hadamard product (i.e., the entrywise product) of $\mathbf{x}$ and $\mathbf{y}$. The notations $\mathbf{A} \succeq \mathbf{B}$ and $\mathbf{A} \succ \mathbf{B}$ mean that $(\mathbf{A} - \mathbf{B})$ is a positive semidefinite and definite matrix, respectively. We also denote $[x]_{+} = \max(x, 0)$.

## II. SYSTEM MODEL

Consider a SU-MIMO channel, where the transmitter is equipped with $N$ antennas and the receiver with $M$ antennas. The channel matrix is represented by $\mathbf{H} \in \mathbb{C}^{M \times N}$, which is assumed to be known perfectly at the transmitter. The received signal is given by

$$\mathbf{y} = \mathbf{H}\mathbf{s} + \mathbf{z} \qquad (1)$$

where $\mathbf{s}$ is the vector of transmitted symbols, and $\mathbf{z} \in \mathbb{C}^{M \times 1}$ is the background noise with distribution $\mathcal{CN}(\mathbf{0}, \mathbf{I}_M)$. Let $\mathbf{S} = E\{\mathbf{s}\mathbf{s}^{\dagger}\}$ be the input covariance matrix for the transmitted signal. We are interested in finding the capacity of the above channel with PAPC, which is formulated as

$$\underset{\mathbf{S} \succeq \mathbf{0}}{\text{maximize}} \quad \log |\mathbf{I} + \mathbf{H}\mathbf{S}\mathbf{H}^{\dagger}| \qquad (2a)$$
$$\text{subject to} \quad [\mathbf{S}]_{i,i} \leq P_i, \ i = 1, 2, \ldots, N \qquad (2b)$$

where $P_i$ is the maximum power constraint on the $i$th antenna. The problem (2) is a convex program, and can be solved by general purpose optimization software.[1] However, the computational complexity of these convex solvers, which are usually based on interior-point methods, increases rapidly with the number of transmit antennas $N$, thereby not suitable for large-scale MIMO systems. Herein, we propose two efficient iterative algorithms which will be numerically shown to achieve a superlinear convergence rate.

## III. FIXED-POINT ITERATION

We first note that the Slater condition is satisfied for (2) and thus strong duality holds. Now, consider the partial Lagrangian function of (2), which is given by

$$\mathcal{L}(\mathbf{S}, \boldsymbol{\Lambda}) = \log|\mathbf{I} + \mathbf{H}\mathbf{S}\mathbf{H}^\dagger| - \text{tr}(\boldsymbol{\Lambda}(\mathbf{S} - \mathbf{P})) \quad (3)$$

where $\mathbf{P} = \text{diag}(P_1, P_2, \ldots, P_N)$, and $\boldsymbol{\Lambda} = \text{diag}(\lambda_1, \lambda_2, \ldots, \lambda_N)$ is the diagonal matrix comprising the dual variables for the $N$ power constraints in (2b). The dual objective of (2) is

$$g(\boldsymbol{\Lambda}) = \max_{\mathbf{S} \succeq \mathbf{0}} \mathcal{L}(\mathbf{S}, \boldsymbol{\Lambda}). \quad (4)$$

To find the optimal solution of (2), we only need to consider the case where $\boldsymbol{\Lambda} \succ \mathbf{0}$, i.e, $\lambda_i > 0$ for all $i$, otherwise $g(\boldsymbol{\Lambda})$ is unbounded above, which cannot be the dual optimal of (2). This can be easily seen by contradiction. Suppose $\lambda_i = 0$ for some $i$. Then create a diagonal matrix $\mathbf{S} = \text{diag}([0, \ldots, 0, \alpha_i, 0, \ldots, 0]^T)$. Accordingly, we can check that $\mathcal{L}(\mathbf{S}, \boldsymbol{\Lambda}) = \log(1 + \alpha_i \sum_{j=1}^M |\mathbf{H}_{j,i}|^2) \to \infty$ if $\alpha_i \to \infty$. Moreover, for a given $\boldsymbol{\Lambda} \succ \mathbf{0}$, we can solve (4) efficiently as described next. Let us denote $\hat{\mathbf{S}} = \boldsymbol{\Lambda}^{1/2}\mathbf{S}\boldsymbol{\Lambda}^{1/2}$. Then finding $\mathbf{S}$ to maximize $\mathcal{L}(\mathbf{S}, \boldsymbol{\Lambda})$ amounts to solving the following problem

$$\begin{aligned} \text{maximize} \quad & \log|\mathbf{I} + \mathbf{H}\boldsymbol{\Lambda}^{-1/2}\hat{\mathbf{S}}\boldsymbol{\Lambda}^{-1/2}\mathbf{H}^\dagger| - \text{tr}(\hat{\mathbf{S}}) \\ \text{subject to} \quad & \hat{\mathbf{S}} \succeq \mathbf{0}. \end{aligned} \quad (5)$$

The above problem admits the solution based on waterfilling algorithm with *fixed water level* [6]. Explicitly, let $\mathbf{V}\boldsymbol{\Sigma}\mathbf{V}^\dagger = \boldsymbol{\Lambda}^{-1/2}\mathbf{H}^\dagger\mathbf{H}\boldsymbol{\Lambda}^{-1/2}$ be the eigenvalue decomposition (EVD) of $\boldsymbol{\Lambda}^{-1/2}\mathbf{H}^\dagger\mathbf{H}\boldsymbol{\Lambda}^{-1/2}$, where $\mathbf{V} \in \mathbb{C}^{N \times N}$ are unitary matrix, and $\boldsymbol{\Sigma} \in \mathbb{C}^{N \times N}$ is a matrix of *(possibly zero) eigenvalues in decreasing order* of $\boldsymbol{\Lambda}^{-1/2}\mathbf{H}^\dagger\mathbf{H}\boldsymbol{\Lambda}^{-1/2}$. Let $r = \text{rank}(\mathbf{H}\boldsymbol{\Lambda}^{-1/2})$, and $\rho_i, i = 1, \ldots r$, be $r$ positive eigenvalues of $\mathbf{H}\boldsymbol{\Lambda}^{-1/2}$. Then, $\hat{\mathbf{S}}$ can be found as

$$\hat{\mathbf{S}} = \mathbf{V}\text{diag}([1 - \frac{1}{\rho_1}]_+, \ldots, [1 - \frac{1}{\rho_r}]_+, \underbrace{0, \ldots, 0}_{N-r})\mathbf{V}^\dagger. \quad (6)$$

Consequently, $\mathbf{S}$ is given by

$$\mathbf{S} = \boldsymbol{\Lambda}^{-1/2}\mathbf{V}\text{diag}([1 - \frac{1}{\rho_1}]_+, \ldots, [1 - \frac{1}{\rho_r}]_+, 0, \ldots, 0)\mathbf{V}^\dagger\boldsymbol{\Lambda}^{-1/2}. \quad (7)$$

---
[1]More specifically, (2) in the current form is in fact a MAXDET program [5] but can be reformulated as an SDP for which dedicated solvers are more available.

As a closer look at (7), let $s$ be the largest number such that $1 - \frac{1}{\rho_s} > 0$. Then, $\mathbf{S}$ is equivalently written as

$$\mathbf{S} = \boldsymbol{\Lambda}^{-1/2}\mathbf{V}\text{diag}(1 - \frac{1}{\rho_1}, \ldots, 1 - \frac{1}{\rho_s}, \underbrace{0, \ldots, 0}_{N-s})\mathbf{V}^\dagger\boldsymbol{\Lambda}^{-1/2}. \quad (8)$$

Since $\mathbf{V}\mathbf{V}^\dagger = \mathbf{I}$, $\mathbf{S}$ is further simplified as

$$\mathbf{S} = \boldsymbol{\Lambda}^{-1} - \boldsymbol{\Lambda}^{-1/2}\mathbf{V}\text{diag}(\frac{1}{\rho_1}, \ldots, \frac{1}{\rho_s}, \underbrace{1, \ldots, 1}_{N-s})\mathbf{V}^\dagger\boldsymbol{\Lambda}^{-1/2}. \quad (9)$$

We can prove that at the optimum, $[\mathbf{S}]_{i,i} = P_i$ for all $i = 1, \ldots, N$. Thus, in order to find optimal $\mathbf{S}$, we need to find $\boldsymbol{\Lambda}$ such that

$$\left[\boldsymbol{\Lambda}^{-1} - \boldsymbol{\Lambda}^{-1/2}\mathbf{V}\text{diag}(\frac{1}{\rho_1}, \ldots, \frac{1}{\rho_s}, 1, \ldots, 1)\mathbf{V}^\dagger\boldsymbol{\Lambda}^{-1/2}\right]_{i,i} = P_i. \quad (10)$$

Since $\boldsymbol{\Lambda}$ is a diagonal matrix, (10) equals to

$$\left(\mathbf{I} - \left[\mathbf{V}\big(\text{diag}(\frac{1}{\rho_1}, \ldots, \frac{1}{\rho_s}, 1, \ldots, 1)\mathbf{V}^\dagger\big)\right]_{i,i}\right)\left[\boldsymbol{\Lambda}^{-1}\right]_{i,i} = P_i. \quad (11)$$

Let $\boldsymbol{\Psi}(\tilde{\boldsymbol{\lambda}}) = \left[\mathbf{V}\big(\text{diag}(\frac{1}{\rho_1}, \ldots, \frac{1}{\rho_s}, \mathbf{1}_{N-s})\mathbf{V}^\dagger\big)\right]$. Then, we can rewrite (11) in the form of a nonlinear system as

$$\tilde{\boldsymbol{\lambda}} - \text{diag}(\boldsymbol{\Psi}(\tilde{\boldsymbol{\lambda}})) \odot \tilde{\boldsymbol{\lambda}} = \mathbf{p} \quad (12)$$

where $\tilde{\boldsymbol{\lambda}} \triangleq [\lambda_1^{-1}, \lambda_2^{-1}, \ldots, \lambda_N^{-1}]^T$, $\mathbf{p} \triangleq [P_1, P_2, \ldots, P_N]^T$. It is easy to see that

$$\left[\boldsymbol{\Psi}(\tilde{\boldsymbol{\lambda}})\right]_{i,i} = \sum_{j=1}^N \tilde{\rho}_j |v_{i,j}|^2 \quad (13)$$

where $\tilde{\rho}_j = \frac{1}{\rho_j} < 1$ for $1 \leq j \leq s$, and $\tilde{\rho}_j = 1$ for $s < j \leq N$. Since $\sum_{j=1}^N |v_{i,j}|^2 = 1$, it holds that $\boldsymbol{\Psi}(\tilde{\boldsymbol{\lambda}}) \prec \mathbf{I}$ for all $\tilde{\boldsymbol{\lambda}} \succ \mathbf{0}$, and (12) is thus well defined. Unfortunately, there is no analytical solution to (12), mostly due to the fact that $\boldsymbol{\Psi}(\tilde{\boldsymbol{\lambda}})$ is a nonlinear function of $\tilde{\boldsymbol{\lambda}}$. However, (12) already suggests a way to find $\tilde{\boldsymbol{\lambda}}$ iteratively as follows

$$\tilde{\boldsymbol{\lambda}}_{n+1} = \mathbf{p} + \text{diag}(\boldsymbol{\Psi}(\tilde{\boldsymbol{\lambda}}_n)) \odot \tilde{\boldsymbol{\lambda}}_n \triangleq \mathfrak{I}(\tilde{\boldsymbol{\lambda}}_n). \quad (14)$$

In fact, (14) is written in a fixed-point iteration form and its convergence is stated in the following lemma.

**Lemma 1.** *The iterations in* (14) *converge to the unique fixed-point of* (12), *thereby solving* (2).

The proof of Lemma 1 is provided in Appendix A. The key is to show that $\mathfrak{I}(\mathbf{x})$ is a standard interference function.

We can see that the fixed-point algorithm based on (14) requires iteratively performing EVD of $\boldsymbol{\Lambda}^{-1/2}\mathbf{H}^\dagger\mathbf{H}\boldsymbol{\Lambda}^{-1/2}$. A simple way is to treat it as a new matrix at each iteration, but this is not computationally efficient. Exploiting the fact that the channel matrix $\mathbf{H}$ remains the same during the whole iterative process, we present a way to compute the EVD of $\boldsymbol{\Lambda}^{-1/2}\mathbf{H}^\dagger\mathbf{H}\boldsymbol{\Lambda}^{-1/2}$ more efficiently. To this end, let $\mathbf{H} = \mathbf{G}\mathbf{R}$, where $\mathbf{G}$ is unitary and $\mathbf{R}$ is upper triangular, be a QR factorization of $\mathbf{H}$. Then we can write $\mathbf{H}\boldsymbol{\Lambda}^{-1/2} = (\mathbf{G}\mathbf{R})\boldsymbol{\Lambda}^{-1/2} =$

$\mathbf{G}(\mathbf{R}\boldsymbol{\Lambda}^{-1/2})$. Since $\boldsymbol{\Lambda}$ is diagonal, $\mathbf{R}\boldsymbol{\Lambda}^{-1/2}$ is also an upper triangular matrix. Now let $\mathbf{R}\boldsymbol{\Lambda}^{-1/2} = \mathbf{U}\bar{\boldsymbol{\Sigma}}\mathbf{V}^\dagger$ be the SVD of $\mathbf{R}\boldsymbol{\Lambda}^{-1/2}$. Then the EVD of $\boldsymbol{\Lambda}^{-1/2}\mathbf{H}^\dagger\mathbf{H}\boldsymbol{\Lambda}^{-1/2}$ is simply given by $\mathbf{V}\bar{\boldsymbol{\Sigma}}^2\mathbf{V}^\dagger$. We remark that SVD computation for an upper triangular matrix is much cheaper than for a full matrix [7, p. 492], which leads to a huge reduction in the computation cost of the proposed algorithm. The proposed algorithm based on fixed-point iteration is outlined in Algorithm 1.

---

**Algorithm 1:** Proposed solution based on fixed-point iteration.

**Input:** $\boldsymbol{\Lambda}_0$ diagonal matrix of positive elements, $\epsilon > 0$.
1 Initialization: Set $n := 0$ and $\tau = 1 + \epsilon$.
2 Perform QR decomposition of $\mathbf{H}$: $\mathbf{H} = \mathbf{G}\mathbf{R}$, where $\mathbf{G}$ is a unitary matrix and $\mathbf{R}$ is an upper triangular matrix.
3 **while** $\tau > \epsilon$ **do**
4      Perform the SVD of $\mathbf{R}\boldsymbol{\Lambda}_n^{-1/2}$: $\mathbf{R}\boldsymbol{\Lambda}_n^{-1/2} = \mathbf{U}_n\bar{\boldsymbol{\Sigma}}_n\mathbf{V}_n^\dagger$, where $\bar{\boldsymbol{\Sigma}}_n$ is diagonal. Let $\rho_i = \sigma_i^2$, $i = 1, \ldots, r$, where $\sigma_i$ is the $i$th non-zero entry of $\bar{\boldsymbol{\Sigma}}_n$ and $r = \text{rank}(\mathbf{R}\boldsymbol{\Lambda}_n^{-1/2})$.
5      $\tilde{\boldsymbol{\Sigma}}_n := \text{diag}([1-\rho_1]_+, \ldots, [1-\rho_r]_+, 0, \ldots, 0)$.
6      $\boldsymbol{\Psi}_n := \mathbf{V}_n(\mathbf{I} - \tilde{\boldsymbol{\Sigma}}_n)\mathbf{V}_n^\dagger$.
7      $\mathbf{S}_n := \boldsymbol{\Lambda}_n^{-1} - \boldsymbol{\Lambda}_n^{-1/2}\boldsymbol{\Psi}_n\boldsymbol{\Lambda}_n^{-1/2}$.
8      $\tau = |\text{tr}(\boldsymbol{\Lambda}_n(\mathbf{S}_n - \mathbf{P}))|$.
9      $\tilde{\boldsymbol{\lambda}}_{n+1} = \mathbf{p} + \text{diag}(\boldsymbol{\Psi}_n) \odot \tilde{\boldsymbol{\lambda}}_n$.
10      $\boldsymbol{\Lambda}_{n+1} = (\text{diag}\,\tilde{\boldsymbol{\lambda}}_{n+1})^{-1}$.
11      $n := n + 1$.
12 **end**

**Output:** $\mathbf{S}_n$.

---

*Remark* 1. To solve (2), the work of [3], [4] proposed two different algorithms for two difference cases: $M \geq N$ and $M < N$. Moreover, these algorithms are dedicated to full-rank channel matrices. In this regard, Algorithm 1 is more universal in the sense that it is applicable to channel matrices of any dimension and rank-deficiency. Another issue of the methods presented in [3], [4] is that a complete analytical proof of their convergence is sill missing. On the contrary, Algorithm 1 is provably convergent from an arbitrary starting point $\tilde{\boldsymbol{\lambda}}_0 > 0$. Moreover, analytical and numerical results demonstrate Algorithm 1 achieves lower complexity, compared to the ones in [3], [4].

## IV. Alternating Optimization

The second proposed iterative method exploits an interesting result from the duality between BC and MAC [8], [9]. In fact it is shown that (2) is equivalent to the following minimax optimization problem [10]

$$\begin{aligned}\min_{\mathbf{Q}\succeq\mathbf{0}}\max_{\bar{\mathbf{S}}\succeq\mathbf{0}}\quad & \log\frac{|\mathbf{Q}+\mathbf{H}^\dagger\bar{\mathbf{S}}\mathbf{H}|}{|\mathbf{Q}|} \triangleq f(\mathbf{Q},\bar{\mathbf{S}})\\ \text{subject to}\quad & \text{tr}(\bar{\mathbf{S}}) \leq P, \text{tr}(\mathbf{Q}\mathbf{P}) \leq P\\ & \mathbf{Q}: \text{diagonal}\end{aligned} \quad (15)$$

where $P \triangleq \sum_{i=1}^{N} P_i$. The relationship between (2) and (15) is stated in the following fact.

**Fact 1.** *Let $\bar{\mathbf{S}}^\star$ and $\mathbf{Q}^\star$ be the optimal solutions to* (15). *Denote $\mathbf{U}\boldsymbol{\Sigma}\mathbf{V}^\dagger$ to be an SVD of $\mathbf{H}(\mathbf{Q}^\star)^{-1/2}$ where $\boldsymbol{\Sigma}$ is square and diagonal. Then, the optimal solution $\mathbf{S}^\star$ to* (2) *can be found as*

$$\mathbf{S}^\star = (\mathbf{Q}^\star)^{-1/2}\mathbf{V}\mathbf{U}^\dagger\bar{\mathbf{S}}^\star\mathbf{U}\mathbf{V}^\dagger(\mathbf{Q}^\star)^{-1/2}. \quad (16)$$

The above result is in fact a special case of the MAC-to-BC transformation presented in [9] when applying to a single user system.

It is trivial to see that the optimality of (15) is not affected if the inequalities are made to be equality. To appreciate the idea behind the second proposed method, let us define $\mathcal{Q} \triangleq \{\mathbf{Q}|\mathbf{Q} : \text{diagonal}, \mathbf{Q} \succeq \mathbf{0}, \text{tr}(\mathbf{Q}\mathbf{P}) = P\}$ and $\mathcal{S} = \{\bar{\mathbf{S}}|\bar{\mathbf{S}} \succeq \mathbf{0}, \text{tr}(\bar{\mathbf{S}}) = P\}$. We note that the sets $\mathcal{Q}$ and $\mathcal{S}$ are compact convex. Now, (15) can be rewritten in an abstract form as

$$\min_{\mathbf{Q}\in\mathcal{Q}}\max_{\bar{\mathbf{S}}\in\mathcal{S}} f(\mathbf{Q},\bar{\mathbf{S}}). \quad (17)$$

We note that $f(\mathbf{Q},\bar{\mathbf{S}})$ is concave with $\bar{\mathbf{S}}$, and convex with $\mathbf{Q}$, and twice differentiable. Thus a saddle point $(\mathbf{Q}^*,\bar{\mathbf{S}}^*)$ exist for (17) and it holds that

$$f(\mathbf{Q}^*,\bar{\mathbf{S}}) \leq f(\mathbf{Q}^*,\bar{\mathbf{S}}^*) \leq f(\mathbf{Q},\bar{\mathbf{S}}^*). \quad (18)$$

We can see that solving (15) boils down to finding a saddle point for (17). In fact, this interpretation was used in the interior-point method proposed in [10]. The minimax formulation in (17) also suggests a way to find a saddle point by alternatively optimizing $\mathbf{Q}$ and $\bar{\mathbf{S}}$. This method was also mentioned in [10] but note that it is not provably convergent.

In the second proposed algorithm, we still capitalize on the idea of AO, but do it in a novel way to ensure strict monotonicity. Suppose at the $n$th iteration, we obtain $\mathbf{Q}_n$. Then, in the next iteration, we find $\bar{\mathbf{S}}_n$ as the solution to the following problem

$$\begin{aligned}\underset{\bar{\mathbf{S}}\succeq\mathbf{0}}{\text{maximize}}\quad & \log|\mathbf{Q}_n + \mathbf{H}^\dagger\bar{\mathbf{S}}\mathbf{H}|\\ \text{subject to}\quad & \text{tr}(\bar{\mathbf{S}}) = P.\end{aligned} \quad (19)$$

It is well known that the above problem admits the solution based on water-filling algorithm [1], [2]. More explicitly, let $\mathbf{U}_n\boldsymbol{\Sigma}_n\mathbf{U}_n^\dagger = \mathbf{H}\mathbf{Q}_n^{-1}\mathbf{H}^\dagger$ be the EVD of $\mathbf{H}\mathbf{Q}_n^{-1}\mathbf{H}^\dagger$, where $\boldsymbol{\Sigma}_n = \text{diag}(\rho_1, \rho_2, \ldots, \rho_r)$ is a matrix of non-negative eigenvalues of $\mathbf{H}\mathbf{Q}_n^{-1}\mathbf{H}^\dagger$, and $r = \text{rank}(\mathbf{H}\mathbf{Q}_n^{-1/2})$. Then, $\bar{\mathbf{S}}_n$ can be found as

$$\bar{\mathbf{S}}_n = \mathbf{U}_n\hat{\boldsymbol{\Sigma}}_n\mathbf{U}_n^\dagger \quad (20)$$

where $\hat{\boldsymbol{\Sigma}}_n = \text{diag}([\mu - \frac{1}{\rho_1}]_+, [\mu - \frac{1}{\rho_2}]_+, \ldots, [\mu - \frac{1}{\rho_r}]_+)$ and $\mu$ is the water-level, which is chosen to satisfy the total power constraint

$$\sum_{i=1}^{r}[\mu - \frac{1}{\rho_i}]_+ = P. \quad (21)$$

Note that $\bar{\mathbf{S}}_n$ in (20) is the *unique* solution to (19). To find $\mathbf{Q}_{n+1}$, we invoke the following inequality, which results from the concavity of the $\log\det$ function,

$$\log|\mathbf{Q} + \mathbf{H}^\dagger\bar{\mathbf{S}}_n\mathbf{H}| \leq \log|\boldsymbol{\Phi}_n| + \text{tr}\Big(\boldsymbol{\Phi}_n^{-1}(\mathbf{Q}-\mathbf{Q}_n)\Big) \quad (22)$$

where $\mathbf{\Phi}_n = \mathbf{Q}_n + \mathbf{H}^\dagger \bar{\mathbf{S}}_n \mathbf{H}$. In the second proposed algorithm, $\mathbf{Q}_{n+1}$ is found to optimize the upper bound of (15), i.e., $\mathbf{Q}_{n+1}$ is the solution to the following problem

$$\begin{aligned}\underset{\mathbf{Q} \succeq \mathbf{0}}{\text{minimize}} \quad & \operatorname{tr}\left(\mathbf{\Phi}_n^{-1} \mathbf{Q}\right) - \log|\mathbf{Q}| \\ \text{subject to} \quad & \operatorname{tr}(\mathbf{QP}) = P \\ & \mathbf{Q} : \text{diagonal.}\end{aligned} \quad (23)$$

Since $\mathbf{Q}$ in (23) is in fact diagonal, i.e., $\mathbf{Q} = \operatorname{diag}(\mathbf{q})$, we can rewrite (23) as

$$\begin{aligned}\underset{\mathbf{q} > 0}{\text{minimize}} \quad & \sum_{i=1}^N \phi_{ni} q_i - \log q_i \\ \text{subject to} \quad & \sum_{i=1}^N P_i q_i = P\end{aligned} \quad (24)$$

where $\phi_{ni} = \left[\mathbf{\Phi}_n^{-1}\right]_{i,i}$. Interestingly, the above problem also has a water-filling-like solution as

$$q_i = \frac{1}{\phi_{ni} + \gamma P_i} \quad (25)$$

where $\gamma \geq 0$ is the solution of the equation

$$\sum_{i=1}^N \frac{P_i}{\phi_{ni} + \gamma P_i} = P. \quad (26)$$

From the definition of $\mathbf{\Phi}_n$, it holds that $\left[\mathbf{\Phi}_n\right]_{i,i} \geq \left[\mathbf{Q}_n\right]_{i,i}$, and thus $\phi_{ni} = \left[\mathbf{\Phi}_n^{-1}\right]_{i,i} \leq \left[\mathbf{Q}_n^{-1}\right]_{i,i}$. As the result, we obtain $\sum_{i=1}^N \frac{P_i}{\phi_i} \geq \operatorname{tr}(\mathbf{Q}_n \mathbf{P}) = P$, where the equality holds since $\mathbf{Q}_n$ is the solution to (23) in the previous iteration. Note that the left hand side of (26) is decreasing with $\gamma$, and thus (26) always has a unique solution, which can be found efficiently, e.g., by the bisection or Newton method. The second proposed algorithm based on AO is summarized in Algorithm 2. The main point of Algorithm 2 is the use of the inequality in (22) to optimize $\mathbf{Q}$ for a given $\bar{\mathbf{S}}$. This step will eliminate the ping-pong effect mentioned above and ensure the objective sequence is strictly decreasing. The convergence proof of Algorithm 2 is provided in Appendix B. Note that the error tolerance $\tau$ in Algorithm 2 is only computed for $n \geq 1$.

---

**Algorithm 2:** Proposed solution base on AO.

**Input:** $\mathbf{Q}_0$ is feasible to $\mathcal{Q}$, and $\epsilon > 0$.
1  Initialize $n := 0$, $\tau := 1 + \epsilon$.
2  **while** $\tau > \epsilon$ **do**
3      Apply water-filling algorithm (i.e., (20) and (21)) to compute $\mathbf{S}_{n+1} = \arg\max_{\mathbf{S} \in \mathcal{S}} \log|\mathbf{Q}_n + \mathbf{H}^\dagger \bar{\mathbf{S}} \mathbf{H}|$.
4      $\tau = |f(\mathbf{Q}_n, \bar{\mathbf{S}}_n) - f(\mathbf{Q}_{n-1}, \bar{\mathbf{S}}_{n-1})|$.
5      $\mathbf{\Phi}_n^{-1} := (\mathbf{Q}_n + \mathbf{H}^\dagger \bar{\mathbf{S}}_n \mathbf{H})^{-1}$.
6      Find $\mathbf{Q}_{n+1} = \arg\min_{\mathbf{Q} \in \mathcal{Q}} \operatorname{tr}(\mathbf{\Phi}_n^{-1} \mathbf{Q}) - \log|\mathbf{Q}|$, using (25) and (26).
7      $n := n + 1$.
8  **end**
**Output:** $\bar{\mathbf{S}}_n$ and use (16) to compute optimal $\mathbf{S}$.

---

We remark that line 3 in Algorithm 2 involves the EVD of $\mathbf{H}\mathbf{Q}_n^{-1}\mathbf{H}^\dagger$, which can be computed similarly as done in Algorithm 1 to reduce the overall complexity. Specifically let $\mathbf{GR} = \mathbf{H}$ be the QR decomposition of $\mathbf{H}$. Next we compute the SVD of the upper triangular matrix $\mathbf{RQ}_n^{-1/2}$ as $\tilde{\mathbf{U}}_n \tilde{\mathbf{\Sigma}}_n \tilde{\mathbf{V}}_n^\dagger = \mathbf{RQ}_n^{-1/2}$. Then the EVD of $\mathbf{HQ}_n^{-1}\mathbf{H}^\dagger$ is simply given by $\mathbf{U}_n \mathbf{\Sigma}_n \mathbf{U}_n^\dagger = \mathbf{HQ}_n^{-1}\mathbf{H}^\dagger$, where $\mathbf{U}_n = \mathbf{G}\tilde{\mathbf{U}}_n$ and $\mathbf{\Sigma}_n = \tilde{\mathbf{\Sigma}}_n^2$. Moreover, we note that $\bar{\mathbf{S}}_n$ needs not be computed explicitly as in (20) for each iteration. The reason is that the diagonal elements of $\mathbf{\Phi}_n^{-1}$ in Line 5 can be found efficiently from the SVD of $\mathbf{RQ}_n^{-1/2}$ as shown in the following.

Using the matrix-inversion lemma, we can write $\mathbf{\Phi}_n^{-1} = \mathbf{Q}_n^{-1/2}(\mathbf{I} + \mathbf{Q}_n^{-1/2}\mathbf{H}^\dagger \bar{\mathbf{S}}_n \mathbf{H} \mathbf{Q}_n^{-1/2})^{-1}\mathbf{Q}_n^{-1/2} = \mathbf{Q}_n^{-1/2}(\mathbf{I} + \tilde{\mathbf{V}}_n \hat{\mathbf{\Sigma}}_n \tilde{\mathbf{V}}_n^\dagger)^{-1}\mathbf{Q}_n^{-1/2}$, where the latter equality holds due to (20). Now let $\dot{\mathbf{\Sigma}}_n$ be the diagonal matrix containing all *strictly positive* entry of $\hat{\mathbf{\Sigma}}_n$, and $\dot{\mathbf{V}}_n$ be the corresponding singular vectors. Then we can write $(\mathbf{I} + \tilde{\mathbf{V}}_n \hat{\mathbf{\Sigma}}_n \tilde{\mathbf{V}}_n^\dagger)^{-1} = (\mathbf{I} + \dot{\mathbf{V}}_n \dot{\mathbf{\Sigma}}_n \dot{\mathbf{V}}_n^\dagger)^{-1} \stackrel{(a)}{=} \mathbf{I} - \dot{\mathbf{V}}_n (\dot{\mathbf{\Sigma}}_n^{-1} + \dot{\mathbf{V}}_n^\dagger \dot{\mathbf{V}}_n)^{-1} \dot{\mathbf{V}}_n^\dagger \stackrel{(b)}{=} \mathbf{I} - \dot{\mathbf{V}}_n (\dot{\mathbf{\Sigma}}_n^{-1} + \mathbf{I})^{-1} \dot{\mathbf{V}}_n^\dagger$, where $(a)$ is due to the matrix inversion lemma, and $(b)$ holds true since $\dot{\mathbf{V}}_n^\dagger \dot{\mathbf{V}}_n = \mathbf{I}$. In summary, we have $\mathbf{\Phi}_n^{-1} = \mathbf{Q}_n^{-1} - \mathbf{Q}_n^{-1/2} \dot{\mathbf{V}}_n (\dot{\mathbf{\Sigma}}_n^{-1} + \mathbf{I})^{-1} \dot{\mathbf{V}}_n^\dagger \mathbf{Q}_n^{-1/2}$. Since $\dot{\mathbf{\Sigma}}_n^{-1} + \mathbf{I}$ is diagonal, its inversion can be computed easily. It is also clear that, to compute $\mathbf{\Phi}_n^{-1}$, what we need is only $\tilde{\mathbf{\Sigma}}_n$ and $\tilde{\mathbf{V}}_n$ from the SVD of $\mathbf{RQ}_n^{-1/2}$.

## V. COMPLEXITY ANALYSIS

In this section, we analyze the complexity of the proposed algorithms in the preceding section, counted as the number of flops. Although flop counting is a crude way to measure the actual computational complexity, it somewhat captures the order of the computation load. In the complexity analysis presented in the following, we only consider the main operations having the most significant complexity and ignore those contributing negligibly to the overall complexity (e.g., subtraction or addition). Note for complex matrices, we simply treat every operation as a complex multiplication which is equal to 6 real flops [11], [12]. First, we assume $M \geq N$ (i.e, more receive than transmit antennas) and apply the relevant results presented in [7] and [11] as follows.

*1) Complexity of Algorithm 1:* Algorithm 1 performs a QR decomposition (cf. line 2) at the first iteration and only $\mathbf{R}$ is needed, which requires $4N^2(3M-N)$ flops. In the subsequent iterations, Algorithm 1 involves an SVD of an upper triangular matrix (line 4), in which only $(\mathbf{\Sigma}, \mathbf{V})$ needs to be computed. This step takes $6(2MN^2 + 11N^3)$ flops. We note that other operations in Algorithm 1 have minor complexity, compared to QR decomposition and SVD, and thus are neglected.

*2) Complexity of Algorithm 2:* To reduce the complexity, Algorithm 2 performs a full QR decomposition in the first iteration, which takes $6(4M^2N - 2MN^2 + \frac{2}{3}N^3)$ flops. Then, the complexity incurred in line 3 of Algorithm 2 is due to finding $(\tilde{\mathbf{\Sigma}}_n, \tilde{\mathbf{V}}_n^\dagger)$ in the SVD of the upper triangular matrix $\mathbf{RQ}_n^{-1/2}$. The flop count of the step is $6(2M^2N + 11N^3)$. The water-filling algorithm to find positive eigenmodes that meet the sum power constraint needs $6(2N^2 + 6N)$ flops. The complexity of line 5 (i.e., computing the diagonal elements of

TABLE I
PER-ITERATION COMPLEXITY COMPARISON

| Algorithms | $M \geq N$ | $M < N$ |
|---|---|---|
| Mode-dropping [3], [4] | $6(4MN^2 + 8N^3)$ | $6(4NM^2 + 8M^3)$ $+12(N-M)^3$ |
| Algorithm 1 | $\mathbf{6(2MN^2 + 11N^3)}$ | $\mathbf{6(2NM^2 + 11M^3)}$ |
| Algorithm 2 | $\mathbf{6(2MN^2 + 11N^3)}$ | $\mathbf{6(2NM^2 + 11M^3)}$ |

$\boldsymbol{\Phi}_n^{-1}$) and that of line 6 are lower compared to the remaining steps and thus can be ignored.

*3) Complexity of the mode-dropping algorithm in [4]:* For comparison purpose we now present the complexity of the so-called mode-dropping algorithm proposed in [3], [4]. Specifically, this method requires an SVD of a *full* $M \times N$ matrix, in the first iteration, for which the flop count is $6(4M^2N + 8MN^2 + 9N^3)$. From the second iteration, the most complex operation of the mode-dropping algorithm is to compute an EVD which requires $6(4MN^2 + 8N^3)$ flops.

Basically, the complexity of the proposed algorithms for the case $N > M$ can be obtained by simply switching $N$ and $M$ in the above analytical expressions. However, for the mode-dropping algorithm, two additional matrix inversions need to be performed, resulting in an increased complexity. The per-iteration complexity comparison (after the first iteration) is summarized in Table I, where the bold text refers to the algorithm with the lowest complexity, i.e., Algorithm 1 and 2. However, the total complexity of an iterative algorithm heavily depends on the number of iterations required to converge. This issue is evaluated for various numerical experiments in Section VI.

## VI. NUMERICAL RESULTS

In this section, we evaluate the performance of our proposed algorithm compared with those in [3], [4]. An error tolerance of $\epsilon = 10^{-6}$ is selected as the stopping criterion for all algorithms. The initial value of $\mathbf{Q}^0$ in Algorithm 2 is set to the identity matrix for all simulations. Other simulation parameters are specified for each setup.

In the first numerical experiment, we plot the convergence rate of Algorithm 1 and the mode-dropping algorithm in [3] for the case $M = N = 3$, and the channel matrix is given by

$$\mathbf{H} = \begin{pmatrix} -0.7936 - 0.3270i & -0.6795 + 0.8721i & -0.1916 - 1.0373i \\ 0.2165 - 0.2898i & -0.4623 + 0.4316i & -0.6364 - 1.1496i \\ -0.8290 - 0.3561i & -0.8693 + 0.0418i & -0.2020 - 1.3893i \end{pmatrix}$$

which is generated randomly. The total power is set to 1 W and the power constraint for the first, second and third antenna are $0.1, 0.45, 0.45$ W, respectively. The initial value of $\mathbf{\Lambda}^0$ in Algorithm 1 is set to the identity matrix. As shown in Fig. 1, Algorithm 2 takes the smallest number of iterations to converge. For Algorithm 1, we can infer from (14) that this algorithm can attain a good convergence rate if all the diagonal entries of $\boldsymbol{\Psi}(\tilde{\boldsymbol{\lambda}}_n)$ are much less than 1, i.e. the singular values of $\mathbf{H}$ and/or $\mathbf{p}$ are relatively large. For the considered channel matrix, the singular values are $2.7976, 0.8656$ and $0.5435$ and the power constraint is relatively small. Therefore, Algorithm

1 converges quite slowly, requiring around 40 iterations. The same observation can be seen in the mode-dropping method.

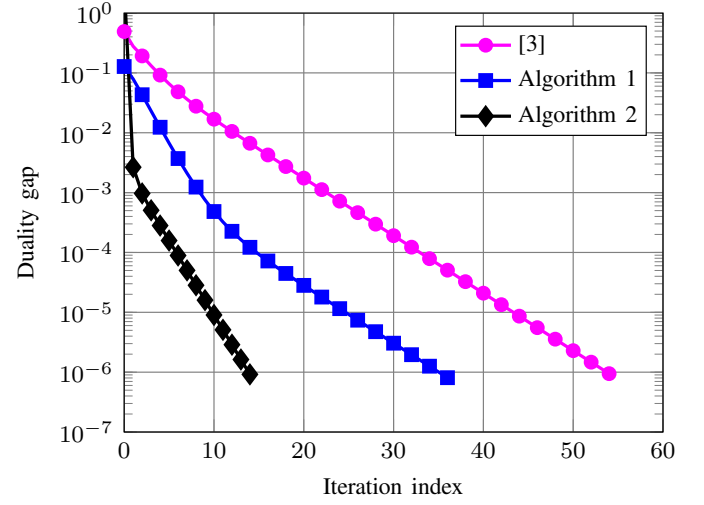

Fig. 1. Convergence comparison for a single user MIMO system with $N = 3$ and $M = 3$.

Fig. 2 plots the average number of iterations required by the iterative algorithms in comparison to converge for the case $N = 4$ and $M = 2$. The results reported in Fig. 2 are averaged among 1000 i.i.d random channel realizations. In particular, the initial value $\mathbf{\Lambda}^0$ for Algorithm 1 is generated the same way as done in [3], [4]. We note that for the case considered in Fig. 2, the second algorithm in [4] is applicable. As can be seen, the convergence rate of Algorithm 2 remains the fastest, which is consistent with the result in Fig. 1. Another observation is that, Algorithm 1 and the method in [4] require the same number of iterations to converge (cf. Fig. 2) as both of them start from the same initial point. In fact, Algorithm 1 can be considered a generalization of algorithms in [3], [4].

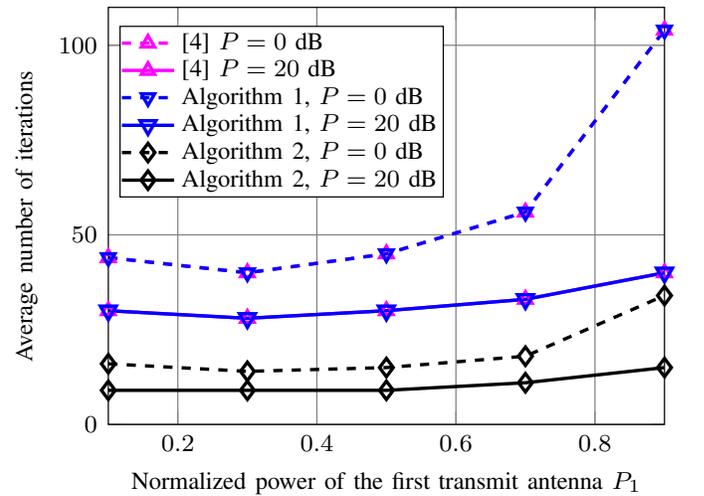

Fig. 2. Average number of iterations required to converge of iterative algorithms with $N = 4$ and $M = 2$.

Finally, we compare the overall complexity of the proposed algorithms with the mode-dropping method in [4]. Specifically, Fig. 3 shows the average number of flops needed by all the iterative algorithms in comparison to converge. In this simulations setting, the number of transmit antennas is set to $N = 30$, and the number of receive antennas is varied from 5 to 25. All transmit antennas are subject to equal power constraint which is equal to $P/N$. The channel realizations are generated the same way as carried out in Fig. 2. As expected, Algorithm 2 achieves the lowest overall complexity since it has not only low per-iteration complexity but also fastest convergence rate. Moreover, the complexity of Algorithm 1 is lower than that of [4], which can be explained by the fact that the per-iteration complexity of Algorithm 1 is lower than that of the mode-dropping algorithm in [4] as mentioned in the preceding section.

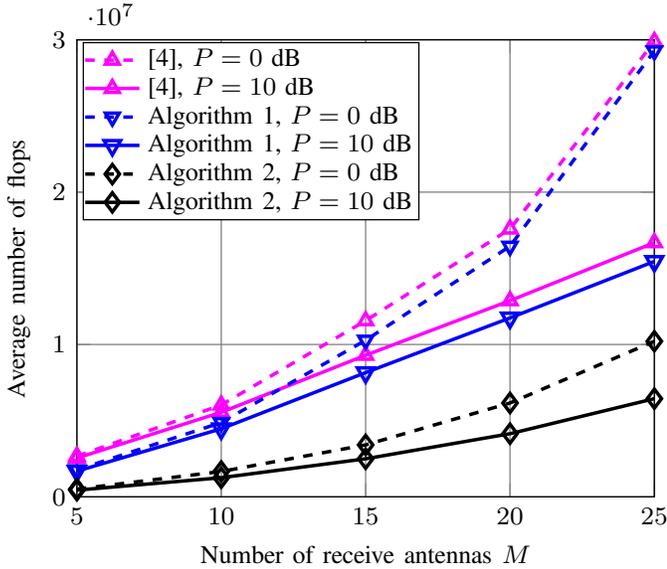

Fig. 3. Total complexity comparison versus the number of receive antennas $M$. The number of transmit antennas is taken as $N = 30$.

## VII. CONCLUSIONS

We have proposed two low-complexity iterative algorithms to compute the capacity of single user MIMO with PAPC. The proposed algorithms are provably convergent. Numerical experiments have been carried out to demonstrate that the proposed algorithms are superior to known solutions in terms of computational complexity. Due to their low-complexity feature, the proposed algorithms can be applied to find the capacity of a massive SU-MIMO system subject to PAPC.

### ACKNOWLEDGEMENTS

This work was supported by a research grant from Science Foundation Ireland and is co-funded by the European Regional Development Fund under Grant 13/RC/2077.

## APPENDIX A
### PROOF OF LEMMA 1

The key to prove the convergence of the fixed-point iteration in (14) is to show that $\mathfrak{I}(\mathbf{x})$ is a standard interference function. That is, for all $\mathbf{x} \geq 0$ then $\mathfrak{I}(\mathbf{x})$ satisfies the following three properties

- Positivity: $\mathfrak{I}(\mathbf{x}) > 0$.
- Monotonicity: If $\mathbf{x} \geq \mathbf{y}$, then $\mathfrak{I}(\mathbf{x}) \geq \mathfrak{I}(\mathbf{y})$.
- Scalability: For all $\alpha > 1, \alpha \mathfrak{I}(\mathbf{x}) > \mathfrak{I}(\alpha \mathbf{x})$.

$$\mathfrak{I}(\alpha \mathbf{x}) = \mathbf{p} + \alpha \operatorname{diag}(\boldsymbol{\Psi}(\alpha \mathbf{x})) \odot \mathbf{x} \quad (27)$$
$$\overset{(a)}{<} \mathbf{p} + \alpha \operatorname{diag}(\boldsymbol{\Psi}(\mathbf{x})) \odot \mathbf{x} \quad (28)$$
$$\overset{(b)}{<} \alpha(\mathbf{p} + \operatorname{diag}(\boldsymbol{\Psi}(\mathbf{x})) \odot \mathbf{x}) = \alpha \mathfrak{I}(\mathbf{x}) \quad (29)$$

where (a) is due to the fact that $\boldsymbol{\Psi}(\alpha \mathbf{x}) < \boldsymbol{\Psi}(\mathbf{x})$ which can be proven from the definition of $\boldsymbol{\Psi}(\mathbf{x})$ as follows. Let $\mathbf{X} = \operatorname{diag}(\mathbf{x})$ and $\mathbf{V}\boldsymbol{\Sigma}\mathbf{V}^\dagger = \mathbf{X}\mathbf{H}^\dagger\mathbf{H}\mathbf{X}$ be the EVD of $\mathbf{X}\mathbf{H}^\dagger\mathbf{H}\mathbf{X}$, where $\boldsymbol{\Sigma} = \operatorname{diag}([\rho_1, \rho_2, \ldots, \rho_r, \mathbf{0}_{N-r}])$ and $r = \operatorname{rank}(\mathbf{H}^\dagger\mathbf{H})$. Then it follows immediately that $\mathbf{V}\tilde{\boldsymbol{\Sigma}}\mathbf{V}^\dagger$ is the EVD of $\tilde{\mathbf{X}}\mathbf{H}^\dagger\mathbf{H}\tilde{\mathbf{X}}$, where $\tilde{\mathbf{X}} = \operatorname{diag}(\alpha \mathbf{x})$, $\tilde{\boldsymbol{\Sigma}} = \operatorname{diag}([\tilde{\rho}_1, \tilde{\rho}_2, \ldots, \tilde{\rho}_r, \mathbf{0}_{N-r}])$, and $\tilde{\rho}_i = \alpha^2 \rho_i$ for $i = 1, 2, \ldots, r$. Thus we have $\frac{1}{\tilde{\rho}_i} = \frac{1}{\alpha^2 \rho_i} < \frac{1}{\rho_i}$ for $\alpha > 1$. Now by the construction of the matrix $\boldsymbol{\Psi}(\cdot)$ from (7) to (11), it is easy to see that $\boldsymbol{\Psi}(\alpha \mathbf{x}) \prec \boldsymbol{\Psi}(\mathbf{x})$ and thus $\operatorname{diag}(\boldsymbol{\Psi}(\alpha \mathbf{x})) < \operatorname{diag}(\boldsymbol{\Psi}(\mathbf{x}))$. The inequality (b) holds since $\mathbf{p} < \alpha \mathbf{p}$ for $\alpha > 1$, which results in $\mathbf{p} + \alpha \operatorname{diag}(\boldsymbol{\Psi}(\mathbf{x})) \odot \mathbf{x} < \alpha(\mathbf{p} + \operatorname{diag}(\boldsymbol{\Psi}(\mathbf{x})) \odot \mathbf{x}) = \alpha \mathfrak{I}(\mathbf{x})$.

To prove the monotonicity of $\mathfrak{I}(\mathbf{x})$, we need to show that for all $\mathbf{x}, \mathbf{y} \geq 0$ then $\mathfrak{I}(\mathbf{x}) \geq \mathfrak{I}(\mathbf{y})$ or equivalently $\operatorname{diag}(\boldsymbol{\Psi}(\mathbf{x})) \odot \mathbf{x} \geq \operatorname{diag}(\boldsymbol{\Psi}(\mathbf{y})) \odot \mathbf{y}$. Let $\mathbf{X} = \operatorname{diag}(\mathbf{x}), \mathbf{Y} = \operatorname{diag}(\mathbf{y})$. Then monotonicity proof is equivalent to showing that $\operatorname{diag}(\mathbf{X}^{1/2}\boldsymbol{\Psi}(\mathbf{x})\mathbf{X}^{1/2}) \geq \operatorname{diag}(\mathbf{Y}^{1/2}\boldsymbol{\Psi}(\mathbf{y})\mathbf{Y}^{1/2})$ for $\mathbf{X} \succeq \mathbf{Y} \succeq \mathbf{0}$.

Let us first consider the case $N \leq M$ and $\mathbf{H}$ is full column rank. Then we can write the EVD of $\boldsymbol{\Lambda}^{-1/2}\mathbf{H}^\dagger\mathbf{H}\boldsymbol{\Lambda}^{-1/2}$ as

$$\underbrace{\boldsymbol{\Lambda}^{-1/2}\mathbf{H}^\dagger\mathbf{H}\boldsymbol{\Lambda}^{-1/2}}_{\mathbf{B}} = \mathbf{V}\boldsymbol{\Sigma}\mathbf{V}^\dagger. \quad (30)$$

For notational convenience, let $\mathbf{K} = \mathbf{H}^\dagger\mathbf{H}$. Note that $\mathbf{K}$ is full-rank and thus invertible. Then the above equation can be rewritten as

$$\mathbf{B}^{-1} = \boldsymbol{\Lambda}^{1/2}\mathbf{K}^{-1}\boldsymbol{\Lambda}^{1/2} = \mathbf{V}\boldsymbol{\Sigma}^{-1}\mathbf{V}^\dagger \quad (31)$$

which the results in

$$\mathbf{B}^{-1} - \mathbf{I} = \mathbf{V}(\boldsymbol{\Sigma}^{-1} - \mathbf{I})\mathbf{V}^\dagger. \quad (32)$$

Let $\tilde{\boldsymbol{\Sigma}}$ be the $(N-k)$ *positive* eigenvalues of $\mathbf{B}^{-1} - \mathbf{I}$ and $\tilde{\mathbf{V}}_k$ consist of the corresponding $N - k$ eigenvectors, $\bar{\boldsymbol{\Sigma}}$ be the $k$ *non-positive* eigenvalues of $\mathbf{B}^{-1} - \mathbf{I}$, and $\bar{\mathbf{V}}_k$ consist of the corresponding $k$ eigenvectors, and define

$$\mathbf{A}^+ = \tilde{\mathbf{V}}_k \tilde{\boldsymbol{\Sigma}} \tilde{\mathbf{V}}_k^\dagger \quad (33a)$$
$$\mathbf{A}^- = \bar{\mathbf{V}}_k \bar{\boldsymbol{\Sigma}} \bar{\mathbf{V}}_k^\dagger. \quad (33b)$$

Then it holds that

$$\mathbf{B}^{-1} - \mathbf{I} = \mathbf{A}^+ + \mathbf{A}^- \quad (34)$$

and that $\mathbf{A}^-\mathbf{A}^+ = \mathbf{0}$. Now we can write $\mathbf{\Psi}(\tilde{\boldsymbol{\lambda}}) = \mathbf{A}^- + \mathbf{I} = \mathbf{B}^{-1} - \mathbf{A}^+$ and thus

$$\begin{aligned}[\mathbf{\Psi}(\tilde{\boldsymbol{\lambda}})\boldsymbol{\Lambda}^{-1}]_{i,i} &= [\boldsymbol{\Lambda}^{-1/2}\mathbf{\Psi}(\tilde{\boldsymbol{\lambda}})\boldsymbol{\Lambda}^{-1/2}]_{i,i}\\ &= [\boldsymbol{\Lambda}^{-1/2}(\mathbf{B}^{-1} - \mathbf{A}^+)\boldsymbol{\Lambda}^{-1/2}]_{i,i}\\ &= [\boldsymbol{\Lambda}^{-1/2}\mathbf{B}^{-1}\boldsymbol{\Lambda}^{-1/2}]_{i,i} - [\boldsymbol{\Lambda}^{-1/2}\mathbf{A}^+\boldsymbol{\Lambda}^{-1/2}]_{i,i}\\ &= [\mathbf{K}^{-1}]_{i,i} - [\boldsymbol{\Lambda}^{-1/2}\mathbf{A}^+\boldsymbol{\Lambda}^{-1/2}]_{i,i}\\ &= [\mathbf{K}^{-1}]_{i,i} - [\tilde{\boldsymbol{\Lambda}}^{1/2}\mathbf{A}^+\tilde{\boldsymbol{\Lambda}}^{1/2}]_{i,i}.\end{aligned} \quad (35)$$

To proceed further we need to show that if $\mathbf{X} \succeq \mathbf{Y}$ then

$$[\mathbf{X}^{1/2}\mathbf{A}_X^+\mathbf{X}^{1/2}]_{i,i} \leq [\mathbf{Y}^{1/2}\mathbf{A}_Y^+\mathbf{Y}^{1/2}]_{i,i}. \quad (36)$$

Now from (34) we have

$$\mathbf{X}^{-1/2}\mathbf{K}^{-1}\mathbf{X}^{-1/2} - \mathbf{I} = \mathbf{A}_X^+ + \mathbf{A}_X^- \quad (37)$$

which is equivalent to

$$\mathbf{K}^{-1} - \mathbf{X} = \mathbf{X}^{1/2}\mathbf{A}_X^+\mathbf{X}^{1/2} + \mathbf{X}^{1/2}\mathbf{A}_X^-\mathbf{X}^{1/2}. \quad (38)$$

The same result applies to $\mathbf{Y}$, i.e.,

$$\mathbf{K}^{-1} - \mathbf{Y} = \mathbf{Y}^{1/2}\mathbf{A}_Y^+\mathbf{Y}^{1/2} + \mathbf{Y}^{1/2}\mathbf{A}_Y^-\mathbf{Y}^{1/2}. \quad (39)$$

Since $\mathbf{X} \succeq \mathbf{Y} \succ \mathbf{0}$ it holds that

$$\mathbf{K}^{-1} - \mathbf{X} \preceq \mathbf{K}^{-1} - \mathbf{Y}. \quad (40)$$

Substituting (38) and (39) into (40) yields

$$\mathbf{X}^{1/2}\mathbf{A}_X^+\mathbf{X}^{1/2} + \mathbf{X}^{1/2}\mathbf{A}_X^-\mathbf{X}^{1/2}\\ \preceq \mathbf{Y}^{1/2}\mathbf{A}_Y^+\mathbf{Y}^{1/2} + \mathbf{Y}^{1/2}\mathbf{A}_Y^-\mathbf{Y}^{1/2}. \quad (41)$$

We now recall the following inequality. For Hermitian matrices $\mathbf{A}$ and $\mathbf{B}$, if $\mathbf{A} \succeq \mathbf{B}$, then $\mathbf{S}\mathbf{A}\mathbf{S}^H \succeq \mathbf{S}\mathbf{B}\mathbf{S}^H$ for $\mathbf{S} \succeq \mathbf{0}$ [14, Observation 7.7.2]. Applying this inequality to (41) leads to

$$\mathbf{A}_X^+ - \mathbf{X}^{-1/2}\mathbf{Y}^{1/2}\mathbf{A}_Y^+\mathbf{Y}^{1/2}\mathbf{X}^{-1/2}\\ \preceq \mathbf{X}^{-1/2}\mathbf{Y}^{1/2}\mathbf{A}_Y^-\mathbf{Y}^{1/2}\mathbf{X}^{-1/2} - \mathbf{A}_X^- \quad (42)$$

which is then equivalent to

$$\begin{aligned}&\left(\mathbf{A}_X^+\right)^{1/2}\!\left(\mathbf{A}_X^+ - \mathbf{X}^{-1/2}\mathbf{Y}^{1/2}\mathbf{A}_Y^+\mathbf{Y}^{1/2}\mathbf{X}^{-1/2}\right)\!\left(\mathbf{A}_X^+\right)^{1/2}\\ &\preceq \left(\mathbf{A}_X^+\right)^{1/2}\!\left(\mathbf{X}^{-1/2}\mathbf{Y}^{1/2}\mathbf{A}_Y^-\mathbf{Y}^{1/2}\mathbf{X}^{-1/2} - \mathbf{A}_X^-\right)\!\left(\mathbf{A}_X^+\right)^{1/2}\\ &\preceq \mathbf{0}.\end{aligned} \quad (43)$$

The above inequality holds true since $\left(\mathbf{A}_X^+\right)^{1/2}\mathbf{A}_X^-\left(\mathbf{A}_X^+\right)^{1/2} = \mathbf{0}$. It is easy to see that (43) results in

$$\mathbf{X}^{1/2}\mathbf{A}_X^+\mathbf{X}^{1/2} \preceq \mathbf{Y}^{1/2}\mathbf{A}_Y^+\mathbf{Y}^{1/2} \quad (44)$$

and thus

$$[\mathbf{X}^{1/2}\mathbf{A}_X^+\mathbf{X}^{1/2}]_{i,i} \leq [\mathbf{Y}^{1/2}\mathbf{A}_Y^+\mathbf{Y}^{1/2}]_{i,i} \quad (45)$$

for all $i$. Here we have used a well-known fact that for $\mathbf{A} \succeq \mathbf{B}$, then $[\mathbf{A}]_{i,i} \geq [\mathbf{B}]_{i,i}$.

We now turn our attention to the general case where $\mathbf{K}$ is singular. This occurs when $N > M$ or $N \leq M$ but $\mathbf{H}$ is not full column rank, i.e. the columns of $\mathbf{H}$ are not linearly independent. First we add a small regularization term to both sides of (30) to obtain

$$\begin{aligned}\mathbf{B}_\epsilon &= \boldsymbol{\Lambda}^{-1/2}\mathbf{H}^\dagger\mathbf{H}\boldsymbol{\Lambda}^{-1/2} + \epsilon\boldsymbol{\Lambda}^{-1}\\ &= \boldsymbol{\Lambda}^{-1/2}\left(\mathbf{H}^\dagger\mathbf{H} + \epsilon\mathbf{I}\right)\boldsymbol{\Lambda}^{-1/2}\\ &= \mathbf{V}\boldsymbol{\Sigma}\mathbf{V}^\dagger + \epsilon\boldsymbol{\Lambda}^{-1}.\end{aligned} \quad (46)$$

We note that $\mathbf{B}_\epsilon$ is invertible for any $\epsilon > 0$. Let $\mathbf{V}_\epsilon\boldsymbol{\Sigma}_\epsilon\mathbf{V}_\epsilon^\dagger$ be the EVD of $\mathbf{B}_\epsilon$ and thus

$$\mathbf{B}_\epsilon^{-1} = \mathbf{V}_\epsilon\boldsymbol{\Sigma}_\epsilon^{-1}\mathbf{V}_\epsilon^\dagger. \quad (47)$$

Applying the result for the nonsingular case, we achieve the following inequality

$$\mathbf{\Psi}_\epsilon(\mathbf{x})\mathbf{X} \succeq \mathbf{\Psi}_\epsilon(\mathbf{y})\mathbf{Y} \quad (48)$$

for arbitrarily small $\epsilon$ and $\mathbf{X} \succeq \mathbf{Y}$, and $\mathbf{\Psi}_\epsilon(\cdot)$ in constructed from $\mathbf{B}_\epsilon$. To complete the proof we are left to show that $\mathbf{\Psi}_\epsilon(\tilde{\boldsymbol{\lambda}})$ is continuous with $\epsilon$, i.e., $\lim_{\epsilon\to 0^+}\mathbf{\Psi}_\epsilon(\tilde{\boldsymbol{\lambda}}) = \mathbf{\Psi}(\tilde{\boldsymbol{\lambda}}) = \mathbf{A}^- + \mathbf{I}$.

To proceed, we note that (34) is changed into

$$\mathbf{B}_\epsilon^{-1} - \mathbf{I} = \mathbf{A}_\epsilon^+ + \mathbf{A}_\epsilon^- \quad (49)$$

where $\mathbf{A}_\epsilon^+$ and $\mathbf{A}_\epsilon^-$ are defined similarly to (33). We will show that $\lim_{\epsilon\to 0}\mathbf{A}_\epsilon^- \to \mathbf{A}^-$. To this end let $\epsilon_{\min} = \epsilon \times \min_i\{1/\lambda_i\}$ and $\epsilon_{\max} = \epsilon \times \max_i\{1/\lambda_i\}$, where $\lambda_i$ is the $i$th diagonal entry of $\boldsymbol{\Lambda}$. It is clear from from (46) that the following inequality holds

$$\underbrace{\mathbf{V}\left((\boldsymbol{\Sigma} + \epsilon_{\min}\mathbf{I})^{-1} - \mathbf{I}\right)\mathbf{V}^\dagger}_{\boldsymbol{\Xi}_{\epsilon_{\min}}} \succ \mathbf{B}_\epsilon^{-1} - \mathbf{I}\\ \succ \underbrace{\mathbf{V}\left((\boldsymbol{\Sigma} + \epsilon_{\max}\mathbf{I})^{-1} - \mathbf{I}\right)\mathbf{V}^\dagger}_{\boldsymbol{\Xi}_{\epsilon_{\max}}}. \quad (50)$$

Further, the matrix $\boldsymbol{\Xi}_{\epsilon_{\min}}$ can be explicitly written as

$$\boldsymbol{\Xi}_{\epsilon_{\min}} = \mathbf{V}\,\mathrm{diag}([\frac{1}{\rho_1 + \epsilon_{\min}} - 1, \ldots, \frac{1}{\rho_r + \epsilon_{\min}} - 1,\\ \underbrace{\frac{1}{\epsilon_{\min}} - 1, \ldots, \frac{1}{\epsilon_{\min}} - 1}_{(N-r)\ \text{terms}}])\mathbf{V}^\dagger \quad (51)$$

where $r = \mathrm{rank}(\mathbf{K})$. Following (34), we decompose $\boldsymbol{\Xi}_{\epsilon_{\min}}$ as

$$\boldsymbol{\Xi}_{\epsilon_{\min}} = \mathbf{A}_{\epsilon_{\min}}^+ + \mathbf{A}_{\epsilon_{\min}}^- \quad (52)$$

where $\mathbf{A}_{\epsilon_{\min}}^+$ and $\mathbf{A}_{\epsilon_{\min}}^-$ consists of positive and non-positive eigenvalues, respectively. As $\epsilon \to 0^+$ we have $\frac{1}{\rho_i + \epsilon_{\min}} \to \frac{1}{\rho_i}$ for all $i = 1, 2, \ldots, r$, and $\frac{1}{\epsilon_{\min}} \gg 1$. Thus, the term $\frac{1}{\epsilon_{\min}} - 1$ in (51) becomes strictly positive and thus is excluded in $\mathbf{A}_{\epsilon_{\min}}^-$. As a result, we have $\lim_{\epsilon\to 0^+}\mathbf{A}_{\epsilon_{\min}}^- = \mathbf{A}^-$. Following the same arguments we can also show that $\lim_{\epsilon\to 0^+}\mathbf{A}_{\epsilon_{\max}}^- = \mathbf{A}^-$. From (50) it is clear that $\lim_{\epsilon\to 0^+}\mathbf{A}_\epsilon^- = \mathbf{A}^-$ and thus

$$\lim_{\epsilon\to 0^+}\mathbf{\Psi}_\epsilon(\tilde{\boldsymbol{\lambda}}) = \lim_{\epsilon\to 0^+}(\mathbf{A}_\epsilon^- + \mathbf{I}) = \mathbf{A}^- + \mathbf{I} = \mathbf{\Psi}(\tilde{\boldsymbol{\lambda}}). \quad (53)$$

By the continuity property shown above, the monotonicity of Algorithm 1 also holds for the singular case, which completes the proof.

## Appendix B
## Convergence Proof of Algorithm 2

We note that the function $\log|\mathbf{Q}+\mathbf{H}^\dagger\bar{\mathbf{S}}\mathbf{H}|$ is *jointly concave* with $\mathbf{Q}$ and $\bar{\mathbf{S}}$. Thus the following inequality holds

$$\log|\mathbf{Q}+\mathbf{H}^\dagger\bar{\mathbf{S}}\mathbf{H}| \leq \log|\underbrace{\mathbf{Q}_n+\mathbf{H}^\dagger\bar{\mathbf{S}}_n\mathbf{H}}_{\mathbf{\Phi}_n}|+\text{tr}(\mathbf{\Phi}_n^{-1}(\mathbf{Q}-\mathbf{Q}_n)) \\ +\text{tr}(\mathbf{H}\mathbf{\Phi}_n^{-1}\mathbf{H}^\dagger(\bar{\mathbf{S}}-\bar{\mathbf{S}}_n)) \quad (54)$$

for all $\mathbf{Q} \in \mathcal{Q}$ and $\bar{\mathbf{S}} \in \mathcal{S}$. The above inequality comes from the first order approximation of $\log|\mathbf{Q}+\mathbf{H}^\dagger\bar{\mathbf{S}}\mathbf{H}|$ around the point $(\mathbf{Q}_n, \bar{\mathbf{S}}_n)$. Substitute $\mathbf{Q} := \mathbf{Q}_{n+1}$ and $\bar{\mathbf{S}} := \bar{\mathbf{S}}_{n+1}$ into the above equality, we have

$$\log|\mathbf{Q}_{n+1}+\mathbf{H}^\dagger\bar{\mathbf{S}}_{n+1}\mathbf{H}| \leq \log|\mathbf{\Phi}_n|+\text{tr}(\mathbf{\Phi}_n^{-1}(\mathbf{Q}_{n+1}-\mathbf{Q}_n)) \\ +\text{tr}(\mathbf{H}\mathbf{\Phi}_n^{-1}\mathbf{H}^\dagger(\bar{\mathbf{S}}_{n+1}-\bar{\mathbf{S}}_n)). \quad (55)$$

Since $\bar{\mathbf{S}}_n = \underset{\bar{\mathbf{S}} \in \mathcal{S}}{\arg\max} \log|\mathbf{Q}_n + \mathbf{H}^\dagger\bar{\mathbf{S}}\mathbf{H}|$, the optimality condition results in

$$\text{tr}(\mathbf{H}\mathbf{\Phi}_n^{-1}\mathbf{H}^\dagger(\bar{\mathbf{S}}-\bar{\mathbf{S}}_n)) \leq 0 \quad (56)$$

for all $\bar{\mathbf{S}} \in \mathcal{S}$. For $\bar{\mathbf{S}} = \bar{\mathbf{S}}_{n+1}$ the above inequality means

$$\text{tr}(\mathbf{H}\mathbf{\Phi}_n^{-1}\mathbf{H}^\dagger(\bar{\mathbf{S}}_{n+1}-\bar{\mathbf{S}}_n)) \leq 0 \quad (57)$$

which leads to

$$\log|\mathbf{Q}_{n+1}+\mathbf{H}^\dagger\bar{\mathbf{S}}_{n+1}\mathbf{H}| \leq \log|\mathbf{\Phi}_n|+\text{tr}(\mathbf{\Phi}_n^{-1}(\mathbf{Q}_{n+1}-\mathbf{Q}_n)). \quad (58)$$

Subtract both sides of the above inequality by $\log|\mathbf{Q}_{n+1}|$ results in

$$f(\mathbf{Q}_{n+1}, \bar{\mathbf{S}}_{n+1}) = \log|\mathbf{Q}_{n+1}+\mathbf{H}^\dagger\bar{\mathbf{S}}_{n+1}\mathbf{H}| - \log|\mathbf{Q}_{n+1}| \\ \leq \log|\mathbf{\Phi}_n| + \text{tr}(\mathbf{\Phi}_n^{-1}(\mathbf{Q}_{n+1}-\mathbf{Q}_n)) - \log|\mathbf{Q}_{n+1}|. \quad (59)$$

Since $\mathbf{Q}_{n+1}$ solves (23) it holds that

$$\log|\mathbf{\Phi}_n| + \text{tr}\big(\mathbf{\Phi}_n^{-1}(\mathbf{Q}_{n+1}-\mathbf{Q}_n)\big) - \log|\mathbf{Q}_{n+1}| \\ \leq \log|\mathbf{\Phi}_n| + \text{tr}\big(\mathbf{\Phi}_n^{-1}(\mathbf{Q}-\mathbf{Q}_n)\big) - \log|\mathbf{Q}| \quad (60)$$

for all $\mathbf{Q} \in \mathcal{Q}$. For the special case $\mathbf{Q} := \mathbf{Q}_n$, the above inequality is reduced to

$$\log|\mathbf{\Phi}_n| + \text{tr}\big(\mathbf{\Phi}_n^{-1}(\mathbf{Q}_{n+1}-\mathbf{Q}_n)\big) - \log|\mathbf{Q}_{n+1}| \\ \leq \underbrace{\log|\mathbf{\Phi}_n| - \log|\mathbf{Q}_n|}_{f(\mathbf{Q}_n, \bar{\mathbf{S}}_n)}. \quad (61)$$

Combining (59) and (61) results in $f(\mathbf{Q}_n, \bar{\mathbf{S}}_n) \geq f(\mathbf{Q}_{n+1}, \bar{\mathbf{S}}_{n+1})$.

It is easy to see that $\{f(\mathbf{Q}_n, \bar{\mathbf{S}}_n)\}$ is bounded above, and thus $\{f(\mathbf{Q}_n, \bar{\mathbf{S}}_n)\}$ is convergent. We also note that (22) is strict if $\mathbf{Q} \neq \mathbf{Q}_n$. Consequently, the sequence $\{f(\mathbf{Q}_n, \bar{\mathbf{S}}_n)\}$ is *strictly decreasing* unless it is convergent. Therefore, the continuity of $f(\cdot)$ and the compactness of $\mathcal{S}$ and $\mathcal{Q}$ imply $\underset{n \to \infty}{\lim} f(\mathbf{Q}_n, \bar{\mathbf{S}}_n) = f(\mathbf{Q}^*, \bar{\mathbf{S}}^*)$.